%% file: main.tex
\documentclass[sigconf]{acmart}
\AtBeginDocument{%
  }

\usepackage{algorithmic}
\usepackage[ruled,linesnumbered]{algorithm2e}
\usepackage{array}
\usepackage{amsfonts}
\usepackage{amssymb}
\usepackage{amsmath}
\usepackage{bm}
\usepackage{booktabs}
\usepackage{color}
\usepackage{epstopdf}
\usepackage{enumitem}
\usepackage{footmisc}
\usepackage{stfloats}  
\usepackage{graphicx}
\usepackage{hyperref}
\usepackage{listings}
\usepackage{multirow}
\usepackage{mathrsfs}
\usepackage{pifont}
\usepackage{subfigure}
\usepackage{setspace}
\usepackage{soul}
\usepackage{textcomp}
\usepackage{url}
\usepackage{xcolor}
\usepackage{xspace}
\usepackage[normalem]{ulem}
\usepackage[np, autolanguage]{numprint}
\usepackage[skip=0pt]{caption}
\usepackage{marvosym}

\usepackage{bbding}

\newcommand{\ie}{\emph{i.e.,}\xspace}

\newcommand{\ignore}[1]{}

\newcommand{\equationref}[1]{Eq~\ref{#1}}
\newcommand{\secref}[1]{Section~\ref{#1}} 
\newcommand{\figref}[1]{Figure~\ref{#1}} 
\newcommand{\tableref}[1]{Table~\ref{#1}} 
\newcommand{\modelname}{\text{Rec-Distill}\xspace}

\begin{document}

\title{Rec-Distill {:} An Industrial Distillation Pipeline for Large-Scale Recommendation Models}

\author{
    Haoran Ding$^{*1}$, Wenlin Zhao$^{*2}$, 
    Yuchen Jiang$^{1}$, Jie Zhu$^{1}$, Juren Li$^{1}$,
    Xinchun Li$^{2}$, Yishujie Zhao$^{2}$, Yi Zhang$^{2}$, 
    Ao Qiao$^{2}$, Jianhui Dong$^{2}$, Cheng Chen$^{2}$, 
    Ziyan Gong$^{1}$, Deping Xie$^{1}$, Peng Xu$^{1}$, 
    Zikai Wang$^{2}$, Yuwei Wang$^{2}$, Huizhi Yang$^{2}$, 
    Zhe Chen$^{1}$, Yuchao Zheng$^{\dagger1}$}

\affiliation{
    \country{ByteDance AML$^{1}$, ByteDance$^{2}$\\
    \{dinghaoran, zhaowenlin, jiangyuchen.jyc, zhujie.zj, lijuren, 
    lixinchun.bu, zhaoyishujie.1, zhangyi.28, qiaoao.516, dongjianhui.167, chencheng.kit, 
    gongziyan, deping.xie, xupeng, 
    wangzikai.kevin, wangyuwei.ah, yanghuizhi,
    chenzhe.john, zhengyuchao.yc\}@bytedance.com}
}

\thanks{* Equal Contributions. \\ $\dagger$ Corresponding authors.}
\renewcommand{\shortauthors}{Haoran Ding et al.}

\begin{abstract}
Large recommendation models have demonstrated substantial potential gains under scaling laws, yet these gains are difficult to realize in industrial recommendation systems because real-world deployment requires lightweight models with strict serving efficiency and latency guarantees. This creates a fundamental gap between offline model scaling and online deployment. 
In this work, we present \textbf{\modelname}, an industrial distillation pipeline that transfers the performance gains of large-scale recommendation modeling to efficient serving models. 
\modelname combines large-teacher scaling with student-side transfer optimization through decoupled training, black-box distillation, debiasing mechanism, and a hybrid batch-streaming pipeline for dynamic recommendation environments. 
Across multiple recommendation and advertising scenarios on real-world platforms, our framework scales teacher models up to 24B dense parameters and 20K behavior sequence length, while enabling lightweight students to recover a substantial portion of teacher gains, with distillation transferability exceeding 60\% in the best setting. 
Extensive offline and online experiments further show that these transferred gains consistently translate into measurable business improvements under industrial constraints. 
These results demonstrate that \modelname provides a practical framework for distilling large-scale recommendation models into deployable, cost-efficient serving systems, while also establishing a reliable path toward scaling recommendation models to even larger regimes in the future.
\end{abstract}



\maketitle

\input{subfile/sec-intro}
\input{subfile/sec-rel}
\input{subfile/sec-pre}
\input{subfile/sec-model}
\input{subfile/sec-exp}
\input{subfile/sec-con}

\balance
\bibliographystyle{ACM-Reference-Format}
\bibliography{main}

\end{document}

%% file: subfile/sec-intro.tex
\section{Introduction}
\label{sec-intro}
In the field of deep learning, the emergence of scaling laws has fundamentally transformed the development trajectory of recommendation systems. 
These laws show that model performance can be predictably and continuously improved by scaling key resources such as the number of parameters, the volume of training data, and computational budget~\cite{kaplan2020scaling,hoffmann2022training}. 
This trend is reshaping the paradigm of recommendation systems by driving researchers to construct increasingly massive models that capture complex, high-order user-item interaction patterns, thereby enabling more accurate personalized recommendations and ultimately improving both user experience and platform revenue~\cite{covington2016deep,zhu2025rankmixer,jiang2026tokenmixerlargescalinglargeranking,chai2025longer}.

However, the continuous expansion of model scale in industrial recommendation systems is bounded by several stringent practical constraints.
First, online services are highly sensitive to response latency. Increased model complexity significantly inflates inference overhead and degrades system responsiveness~\cite{zhou2018deep}.
Second, training cost and return on investment (ROI) impose another critical limitation: the computational requirements for training and deploying ultra-large-scale models grow rapidly, whereas the resulting improvements in online metrics often exhibit diminishing marginal returns, ultimately reducing overall ROI~\cite{sharir2020cost}.

Knowledge Distillation (KD) offers a viable pathway to circumvent this ``scale-efficiency'' dilemma.
Its core idea is to transfer knowledge from a massive teacher model to a lightweight student model, allowing the student to approximate the teacher's performance while substantially reducing inference cost and latency~\cite{hinton2015distilling}. 
Although distillation has achieved remarkable success across various domains~\cite{gou2021knowledge,sanh2019distilbert}, its effective application in dynamic, high-concurrency industrial recommendation scenarios—especially with ultra-large-scale teacher models—still faces salient theoretical and practical limitations.

In CV and NLP domains, KD typically uses a pre-trained teacher to perform a single inference pass over the distillation dataset, or co-trains the student with a frozen teacher~\cite{gou2021knowledge,sanh2019distilbert}. However, this paradigm is fundamentally incompatible with recommendation systems. 
A core requirement of recommendation systems is the ability to continuously learn from streaming data distributions in real time to maintain strong online performance~\cite{khani2024bridging,recommendation2025external}. Once the teacher stops learning from streaming data, its performance degrades rapidly and can no longer effectively guide the student. 
Conversely, co-training requires teacher inference at every student training step, leading to prohibitive training costs that cannot satisfy the rapid iteration requirements of industrial environments.

Furthermore, theoretical studies on distillation indicate that an excessively large capacity gap between teacher and student can impede knowledge transfer and even lead to an inflection point in performance gains~\cite{zhu2022teach, busbridge2025distillation}. 
Existing remedies, such as dynamic teacher-student frameworks or teacher assistants~\cite{guo2024gap,mirzadeh2020teacher}, inevitably introduce additional structural complexity and training overhead. At the same time, the strict industrial requirements for deployment robustness and real-time data processing make many idealized methods—such as those relying on static datasets, white-box access to teacher parameters, or complex feature alignment—impractical for real-world deployment.

Recently, Meta's ExFM framework explored large-scale model serving via external distillation~\cite{recommendation2025external}. However, its primary focus is computational cost amortization, leaving the core question—how to maximize knowledge transfer while scaling the teacher to the extreme—systematically unresolved.

In summary, building a large-scale distillation framework for industrial recommendation requires addressing several core challenges:
\begin{itemize}[topsep=0.5pt, leftmargin=*]
    \item \textbf{Compatibility with multi-stage training pipelines}: 
    Industrial recommendation systems typically rely on multi-stage training paradigms. A practical distillation framework must therefore support reliable writing and reading of distillation signals across different training stages, while keeping training costs manageable and avoiding expensive schemes such as co-training.
    \item \textbf{Support for continual streaming updates}: 
    The framework should support streaming learning for both Teacher and Student. The Teacher should be continuously updated to follow the latest data distribution, while the Student should be trained on both distillation signals and streaming data with minimal latency.
    \item \textbf{Maximizing distillation transferability}: 
    The goal of distillation is to preserve as much as possible of the gains induced by scaling up the Teacher. We characterize this objective as Distillation Transferability. Optimizing transferability requires coordinated design of the distillation objective, the student model, and the underlying data pipeline.
\end{itemize}

To address these challenges, we propose \modelname, a decoupled large-scale distillation framework for industrial recommendation systems. Instead of tightly coupling teacher and student optimization, \modelname separates their training processes, making it possible to scale up the teacher without increasing online inference cost. The teacher continuously tracks the latest data distribution through streaming training, while the distillation signals generated during its forward passes are cached in intermediate storage for student training. This design also enables a flexible ``1-to-N'' paradigm, allowing one teacher to supervise multiple students, even across different stages of the recommendation pipeline (e.g., distilling pre-ranking models from final-ranking teachers)~\cite{zhao2025hybrid}.

To better analyze the distillation problem, we decompose the distillation gain into the product of \textbf{Teacher Gain ($\Delta \text{Gain}_{\text{scale}}$)} and \textbf{Distillation Transferability ($\eta$)}:
\begin{equation}
    \Delta \text{Gain}_{\text{distill}} = \Delta \text{Gain}_{\text{scale}} \times \eta
    \label{eq:gain_distill}
\end{equation}
We define $\Delta \text{Gain}_{\text{scale}} = P_T - P_S^{\text{raw}}$ as the Teacher Gain, and $\eta = (P_S^{\text{distill}} - P_S^{\text{raw}}) / (P_T - P_S^{\text{raw}})$ as the Distillation Transferability, which quantifies the proportion of the teacher's performance advantage successfully absorbed by the student, where $P_T$ denotes the performance of the teacher model, $P_S^{\text{raw}}$ the performance of the student model without distillation, and $P_S^{\text{distill}}$ the performance of the distilled student model.

Therefore, our objective is to improve the overall $\Delta \text{Gain}_{\text{distill}}$ by jointly maximizing both $\Delta \text{Gain}_{\text{scale}}$ and $\eta$.

Since $\Delta \text{Gain}_{\text{scale}} = P_T - P_S^{\text{raw}}$, enhancing teacher gain is equivalent to improving the absolute performance $P_T$ of the teacher model. To maximize $\Delta \text{Gain}_{\text{scale}}$, guided by scaling laws, we comprehensively scale up the teacher model across multiple dimensions. For recommendation systems, this primarily involves three dimensions: dense parameter size, sequence modeling, and training data volume.
 
On the other hand, to maximize Distillation Transferability $\eta$, we optimize from two aspects: the distillation algorithm and the student model architecture. For the distillation algorithm, we adopt a black-box strategy that directly uses the teacher's output logits to optimize the student via cross-entropy loss, keeping the method universal, simple, and efficient. For student model design, we propose a ``decoupled-tower'' architecture, which splits the student network into a main task tower for online inference and an auxiliary tower dedicated to receiving distillation signals. This design not only improves performance through distillation learning but also achieves robust risk isolation for online services.

We summarize the main contributions of this paper as follows:
\begin{itemize}[topsep=0.5pt, leftmargin=*]
    \item \textbf{Industry-leading large-scale distillation paradigm}: 
    To our knowledge, \modelname is the first framework in industrial recommendation scenarios to successfully and efficiently distill a teacher model with up to 24B dense parameters and a 20K sequence length. It systematically validates the efficacy of scaling laws in recommendation distillation and provides a reusable solution for continuous model scaling, effectively decoupling model size from real-world inference cost.
    \item \textbf{Systematic design for high distillation transferability}: 
    Through careful optimization of the distillation algorithm, the innovative ``decoupled-tower'' student architecture, and a unified batch-streaming distillation pipeline, we systematically address the challenges of real-time processing and risk mitigation in industrial distillation. This significantly improves both the robustness of distillation and the efficiency of knowledge transfer.
    \item \textbf{Superior universality and efficacy}: 
    We systematically validate the effectiveness of \modelname. Extensive offline and online experiments across multiple industrial recommendation and advertising scenarios demonstrate that our framework stably achieves peak distillation transferability exceeding 60\%, confirming its broad applicability and high efficiency across diverse business contexts.
\end{itemize}

%% file: subfile/sec-rel.tex
\section{Related Work}
\label{sec-rel}
Knowledge Distillation (KD), introduced by Hinton et al.~\cite{hinton2015distilling}, transfers knowledge from large teacher models to smaller student models by minimizing the difference between their output distributions. 
This approach has been widely used across fields like image classification~\cite{gou2021knowledge} and NLP~\cite{sanh2019distilbert}.
In recommendation systems, KD helps reduce the complexity of large-scale models while maintaining competitive performance~\cite{tang2018ranking,kang2020derd,lee2022collaborative}. 
However, applying KD to recommendation systems presents unique challenges, particularly when scaling up teacher models. 
The large performance gap between teacher and student models, especially in industrial scenarios with ultra-large models and long sequences, can hinder knowledge transfer, resulting in diminishing returns from distillation~\cite{zhu2022teach,busbridge2025distillation}. 
Moreover, most existing methods struggle with the dynamic nature of real-time recommendation systems, where the teacher model is constantly updated, and distillation must be performed efficiently in high-concurrency environments~\cite{khani2024bridging,recommendation2025external}.

Recent related works in scaling KD for recommendation systems include Ranking Distillation~\cite{tang2018ranking}, an early method for distilling ranking models into compact forms, though limited to small scales without incorporating scaling laws or long-sequence modeling; 
DE-RRD~\cite{kang2020derd}, which distills embeddings and ranking distributions to boost student performance but relies on white-box access and neglects batch-streaming engineering or debiasing for industry; 
Unbiased Knowledge Distillation~\cite{chen2023unbiased}, which mitigates selection and popularity biases for improved fairness and accuracy, yet does not scale teachers to billions of parameters, handle extended sequences, or innovate student structures; 
and a recent exploration of KD challenges in online ranking~\cite{khani2024bridging}, emphasizing data shift mitigation and efficient inference in large-scale video recommendation.
The external distillation framework for recommendation systems proposed in~\cite{recommendation2025external} attempted to offload knowledge transfer to external models, reducing the computational burden during inference, but it still does not fully address the challenges of large-scale model distillation in recommendation systems.

Our approach differs by systematically scaling the teacher model (up to 24B parameters) with long sequences (up to 20K tokens) and optimizing distillation for real-time environments. 
We propose a robust student model and distillation pipeline that maximize distillation transferability and allow for efficient deployment in industrial recommendation systems.

%% file: subfile/sec-pre.tex
\section{Preliminaries}
\label{sec-pre}

\subsection{Problem Formulation}
We consider a generalized problem setting in industrial recommendation and advertising systems. \modelname is not restricted to a single prediction task; rather, it provides a unified large-scale distillation framework for diverse business objectives, including but not limited to Click-Through Rate (CTR), Conversion Rate (CVR), and watch-time prediction. Let $\mathcal{U}$ and $\mathcal{I}$ denote the sets of users and items, respectively. For any user-item pair $(u, i)$, the model receives a feature vector $\mathbf{x}_{ui} \in \mathcal{X}$ as input, which typically includes user profile features, item attributes, contextual information, and the user's historical interaction sequence. The objective is to learn a prediction function $f: \mathcal{X} \rightarrow \mathcal{Y}$ to model the user's target feedback $y_{ui}$ on the item. Depending on the task, $y_{ui}$ may be either a discrete behavioral label or a continuous value signal; accordingly, the model is trained with a supervised objective tailored to the task format.

\subsection{Knowledge Distillation for Recommendation}
Knowledge Distillation (KD) aims to transfer the knowledge embedded in a large, high-capacity \textit{teacher model} $f_T$ to a smaller, more efficient \textit{student model} $f_S$.
In the standard KD formulation~\cite{hinton2015distilling}, the student is trained to mimic both the ground-truth labels and the softened output distribution of the teacher. 
Let $z_T = f_T(\mathbf{x})$ and $z_S = f_S(\mathbf{x})$ denote the pre-sigmoid outputs of the teacher and student, respectively. 
The softened probability distributions are obtained by applying a temperature-scaled sigmoid: $p_T = \sigma(z_T/\tau)$ and $p_S = \sigma(z_S/\tau)$, where $\sigma(\cdot)$ is the sigmoid function and $\tau > 0$ is a temperature parameter controlling the smoothness of the distribution. 
The distillation loss is typically defined as the Kullback-Leibler (KL) divergence between these distributions, scaled by $\tau^2$:
\begin{equation}
    \mathcal{L}_{distill} = \tau^2 \cdot D_{KL}(\mathbf{p}_T \| \mathbf{p}_S).
    \label{eq:L_KD}
\end{equation}
The overall student objective is a weighted combination of this distillation loss and the standard task-specific loss $\mathcal{L}_{task}$:
\begin{equation}
    \mathcal{L}_{S} = \alpha \cdot \mathcal{L}_{task} + (1 - \alpha) \cdot \mathcal{L}_{distill},
    \label{eq:L_S}
\end{equation}
where $\alpha \in [0, 1]$ is a balancing weight. 
This formulation aligns with the black-box distillation approach we adopt, in which the student learns exclusively from the teacher's final outputs, without access to its internal representations.

\subsection{Scaling Laws and Distillation Gain}
As established by prior work~\cite{kaplan2020scaling,hoffmann2022training}, scaling laws indicate that model performance $P$ scales predictably as a power-law function of key resources, including model parameter count $N$, training dataset size $D$, and computational budget $C$. In recommendation systems, scaling up $N$ (model capacity) and effectively leveraging longer user behavior sequences (a facet of $D$) are the primary avenues for improving the teacher model's performance $P_T$.

To quantitatively analyze and optimize distillation, we adopt the gain decomposition framework introduced in the Introduction and rewrite \equationref{eq:gain_distill} as follows:
\begin{equation} 
    \eta(t) = \frac{\Delta \text{Gain}_{\text{distill}}(t)}{\Delta \text{Gain}_{\text{scale}}(t)} = \frac{P_S^{\text{distill}}(t) - P_S^{\text{raw}}(t)}{P_T(t) - P_s^{\text{raw}}(t)} 
    \label{eq:eta}
\end{equation}

where $P_S^{\text{raw}}(t)$ denotes the baseline performance of the raw student model without distillation at time $t$, and $P_T(t)$ denotes the teacher's performance at time $t$. Since $P_S^{\text{raw}}(t)$ typically serves as a baseline, $P_T(t)$ becomes a key factor in amplifying distillation gain. The overall distillation gain $\Delta \text{Gain}_{\text{distill}}$ is thus decomposed into the gain from scaling up the teacher, $\Delta \text{Gain}_{\text{scale}}$, and the distillation transferability, $\eta$. Our approach aims to maximize both components through systematic teacher scaling and careful distillation framework design.

%% file: subfile/sec-model.tex
\section{Methodology}
\label{sec-model}

\begin{figure*}[hptb]
    \centering
    \includegraphics[width=0.8\textwidth]{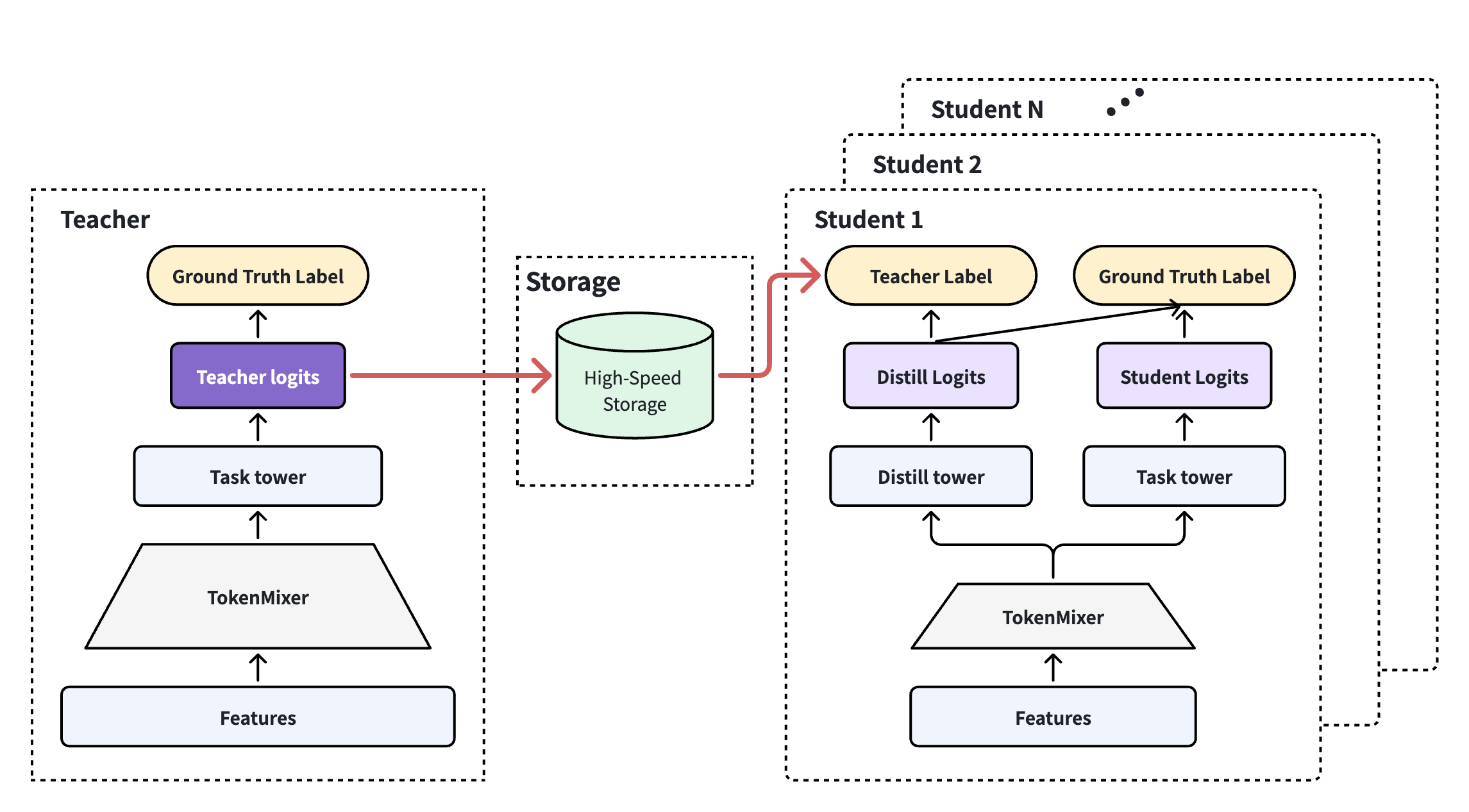}
    \caption{The overall architecture of the \modelname framework. 
    The left side illustrates teacher scaling to maximize $\Delta \text{Gain}_{\text{scale}}$. 
    The right side depicts the student model with its decoupled-tower structure, designed to maximize distillation transferability $\eta$.}
    \label{fig:framework}
\end{figure*}

\subsection{Overall Framework}
The overall architecture of the \modelname distillation pipeline is illustrated in \figref{fig:framework}.

We adopt a \textit{decoupled distillation} paradigm for training the teacher model $f_T$ and the student model $f_S$. 
This framework actualizes a ``1-to-N'' training paradigm, allowing a single teacher to provide supervision signals to multiple independent student models, significantly reducing the training cost per student while naturally adapting to the streaming updates of recommendation systems~\cite{recommendation2025external}.

Specifically, a high-capacity teacher model $f_T$ is trained on the full dataset using ground-truth labels $y$. During forward propagation, the computed logits $p_T$ are persisted into external storage as distillation signals. 
Notably, due to the sparse features in recommendation systems and their strong memorization effects, the conventional ``train-then-predict'' paradigm may introduce data leakage. Therefore, we adopt a forward distillation scheme, in which the teacher's predictions are generated and stored during forward propagation, and then consumed by the Student as distillation targets.

The lightweight student model $f_S$ is built with a decoupled tower design. 
For efficiency, it can be trained on a sampled subset of the data to optimize two components: 1) \textit{the Main Task Tower}, which is responsible for final recommendation prediction during online serving and is optimized exclusively using ground-truth labels; 2) \textit{the Auxiliary Tower}, dedicated strictly to distillation, which is optimized using both ground-truth labels and distillation signals.
The shared backbone parameters in the student model $f_S$ receive gradient updates from both towers, ensuring that the learned representations benefit from both direct task supervision and distilled knowledge.

In industrial environments, this decoupled paradigm also allows both modules to iterate asynchronously, thereby maximizing development efficiency. In particular, the Teacher can continue scaling up without being constrained by online serving cost. At the same time, the Student remains free to evolve according to deployment needs, without being tightly restricted by the Teacher, such as when adding new features or modifying its architecture.

As analyzed previously, optimizing distillation gain is decoupled into two modules: scaling up the teacher (left side of \figref{fig:framework}) to improve \textbf{$\Delta \text{Gain}_{\text{scale}}$}, and refining the student architecture and pipeline (right side) to maximize \textbf{$\eta$}. 

\subsection{Enhancing Teacher Model Gains ($\Delta \text{Gain}_{\text{scale}}$)}
To maximize $\Delta \text{Gain}_{\text{scale}}$, we comprehensively scale up the teacher model. For recommendation systems, scaling primarily involves three dimensions: dense parameter size, sequence modeling, and training data volume.

\begin{itemize}[topsep=0.5pt, leftmargin=*]
    \item \textbf{Dense Parameter Size}: 
    To increase model capacity and representation power, we scale up the number of dense parameters. We adopt the TokenMixer-Large architecture~\cite{jiang2026tokenmixerlargescalinglargeranking}, which employs efficient token-mixing operations as a lightweight alternative to the standard Transformer attention mechanism. This design maintains computational tractability during scaling, enabling us to expand the dense parameter size of the teacher model up to 24B.
    \item \textbf{Sequence Modeling}: 
    Extended user behavior sequences provide richer historical context for modeling long-term interests. To leverage ultra-long sequences, we employ the LONGER architecture~\cite{chai2025longer}. By integrating sequence compression and mixed attention, LONGER circumvents the quadratic complexity bottleneck of standard Transformers, achieving efficient end-to-end modeling for sequences up to 20K in length. This design complements the efficiency of TokenMixer.
    \item \textbf{Training Data Volume}: 
    Drawing insights from the Chinchilla scaling laws~\cite{hoffmann2022training}, which emphasize proportional scaling of model parameters and training data, we augment the data volume used to train the teacher model. A pivotal finding in our experiments is that the teacher can achieve greater gains through data scale-up, and these gains can be effectively transferred via distillation to a student model that requires significantly less training data. This means the student can inherit the benefits of larger datasets without incurring the intrinsic cost of processing them, thereby further amortizing training expense and improving overall ROI.
\end{itemize}

\subsection{Enhancing Distillation Transferability ($\eta$)}
From the definition of distillation transferability in \equationref{eq:eta}, we derive
\begin{equation}
    \eta \propto \frac{P_S^{\text{distill}}}{P_T}.
\end{equation}
It follows that maximizing $\eta$ requires the distilled student's performance $P_S^{\text{distill}}$ to be as close as possible to the teacher's performance $P_T$.
We improve $\eta$ through a novel student architecture with debias mechanism and optimized distillation algorithm design.

\subsubsection{Student Model Design}
The student model in an industrial system has a dual role: it must \textit{learn} effectively from distillation signals during training, and simultaneously provide stable, reliable \textit{serving} during online inference.
To reconcile these needs and enhance system robustness, we propose a novel ``decoupled-tower'' architecture.

The student network is split into two independent towers that share a common feature-processing backbone:

\begin{itemize}[topsep=0.5pt, leftmargin=*]
    \item \textbf{Main Task Tower:} 
    Responsible for generating the final target prediction during online serving. 
    It is trained \textit{exclusively} with standard task loss using ground-truth labels:
    \begin{equation}
        \mathcal{L}_{main} = \mathcal{L}_{task}     
    \end{equation}
    Its parameters are isolated from the distillation process.
    \item \textbf{Auxiliary Tower:} 
    Dedicated to the knowledge distillation process, trained using the combined loss:
    \begin{equation}
        \mathcal{L}_{aux} = \mathcal{L}_S = \mathcal{L}_{task} + \alpha\mathcal{L}_{distill}
    \end{equation}
    It receives the same input features as the main tower and learns to align with the teacher's output distribution.
\end{itemize}

This structural design serves as a critical \textit{fault-tolerance mechanism} for industrial recommendation systems, thereby improving overall system robustness. 
Even when teacher updates are delayed, the distillation pipeline lags behind, or teacher signals become temporarily unstable, the Main Task Tower remains unaffected in both performance and online serving. 
In this way, the decoupled-tower design provides effective risk isolation and ensures stable production serving.


\subsubsection{Student Loss Design}
We employ a black-box distillation strategy, wherein the student is supervised exclusively by the teacher's final output logits. Although white-box methods utilizing the teacher's intermediate features can theoretically provide richer signals, they incur prohibitive communication and storage overheads. 
Thus, black-box distillation offers the best trade-off between efficacy and efficiency.

Numerous prior works demonstrate that different distillation loss designs significantly impact distillation efficacy, with mean squared error (MSE) and cross-entropy (CE) losses being the most common~\cite{gou2021knowledge, yuan2020revisiting}.
Given that our task is binary classification (e.g., CVR prediction), CE loss is naturally suitable. 
Although the labels for student training are soft lables (real numbers), the training objective is to align the student's logit distribution with the teacher's, typically using distribution measures such as \equationref{eq:L_KD}. 
When the teacher distribution is fixed, the KL divergence equates to the CE loss \ie
\begin{equation}
    \mathcal{L}_{distill} = -p_T\log(p_S) - (1-p_T)\log(1-p_S)
    \label{eq:L_distill}
\end{equation}
Thus, setting the temperature parameter $\tau$ to 1 in \equationref{eq:L_KD} makes it equivalent to \equationref{eq:L_distill}.
Theoretically, combining with the preliminaries, the gradient for MSE loss is 
\begin{equation}
    \frac{\partial\mathcal{L}_{MSE}}{\partial \theta} = (p_S - p_T) \cdot p_S (1 - p_S) \frac{\partial z_S}{\partial \theta}
\end{equation}
where $\theta$ denotes model parameters. 
In contrast, the CE loss gradient is
\begin{equation}
    \frac{\partial \mathcal{L}_{CE}}{\partial \theta}=(p_S - p_T) \frac{\partial z_S}{\partial \theta}
\end{equation}
The CE loss exhibits higher gradient efficiency and avoids vanishing gradients (e.g., when $p_S$ approaches 0 or 1, the MSE gradient nears 0).

Therefore, the overall training objective in \equationref{eq:L_S} is expressed in our work as
\begin{equation}
    \mathcal{L}_S = \mathcal{L}_{task} + \alpha\mathcal{L}_{distill}
\end{equation}

In practice, we abandon the conventional constraint that the weights of $\mathcal{L}_{task}$ and $\mathcal{L}_{distill}$ must sum to 1. Empirical observations reveal that the distillation transferability is highly sensitive to the scaling ratio between the distillation loss and the ground-truth task loss. In binary classification scenarios, the scale of $\mathcal{L}_{distill}$ can be up to two orders of magnitude smaller than $\mathcal{L}_{task}$, which severely undermines the distillation efficacy. 
Therefore, by relaxing this constraint and empirically tuning $\alpha$ to align both losses to a comparable order of magnitude, we achieve the optimal distillation performance.

\subsubsection{Student Debias Design}
In industrial recommendation scenarios such as advertising, training data are commonly generated through non-uniform sampling, which introduces systematic bias into the observed label distribution. 

This challenge is particularly important in \modelname, where the teacher and student may be trained on different sample sets. 
The teacher can adopt a more aggressive sampling strategy to enlarge its training set to achieve a better performance. 
However, this also creates a distribution mismatch for the student, which is trained on a smaller training set due to stricter efficiency and training constraints. 
Therefore, beyond standard label correction, the student requires a finer-grained debias mechanism to compensate for the sampling bias induced by heterogeneous teacher-student training data. 

To this end, we apply a sampling-aware correction to transform the raw logit \(z\) into a less biased posterior estimate:

\begin{equation}
    \hat{y}
    =\frac{1}{1+\frac{r_s}{p_X}\left(e^{-z} + 1 - r_{+} + b_s\right)},
    \label{eq:debias_correction}
\end{equation}
where \(r_s\) denotes the negative sampling ratio, \(r_{+}\) denotes the nominal retention rate of positive samples, \(p_X\) denotes the effective retention probability of the target positive event under the current task and traffic condition, and \(b_s\) is an additive bias term introduced by the sampling strategy. We distinguish \(r_{+}\) from \(p_X\) because, in our production pipeline, whether a positive sample is retained may depend not only on the task itself but also on the upstream traffic source. Therefore, while \(r_{+}\) characterizes the global positive-sampling policy, \(p_X\) serves as the effective correction factor used to recover the target probability for each task and traffic pattern.

We denote the above debias correction as a function \(f(\cdot)\). Accordingly, for the teacher model, we distinguish the raw pre-correction logit \(T_1\) from the debiased output \(T_2=f_T(T_1)\), where \(f_T(\cdot)\) is determined by the teacher-side sampling configuration. Similarly, for the student model, we denote its raw logit by \(S_1\) and its debiased output by \(S_2=f_S(S_1)\), where \(f_S(\cdot)\) is defined by the student-side sampling configuration.

\begin{figure*}[hbpt]
    \centering
    \includegraphics[width=0.8\linewidth]{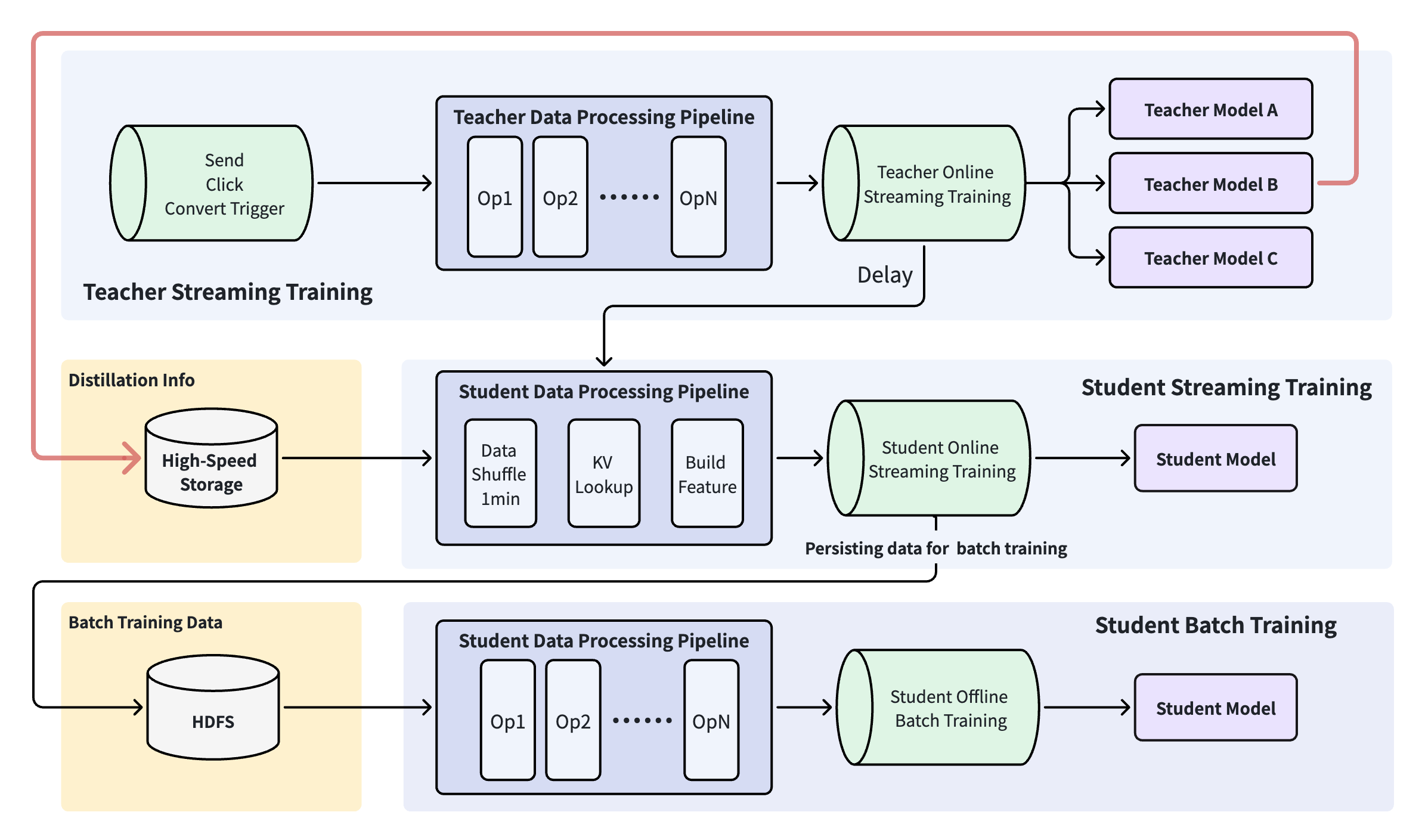}
    \caption{The hybrid batch-streaming distillation pipeline. 
    Batch distillation enables fast initial convergence, while streaming distillation ensures continuous adaptation to real-time data distribution shifts.}
    \label{fig:stream}
\end{figure*}

A key issue in our setting is that the teacher and student are often trained under different sampling strategies, and therefore \(f_T(\cdot)\neq f_S(\cdot)\). As a result, the debiased teacher output \(T_2\) and the debiased student output \(S_2\) are generally not aligned in the same corrected probability space. Directly matching \(T_2\) and \(S_2\) would thus introduce additional bias caused by the mismatch between teacher-side and student-side correction functions.

To address this issue, we first project the teacher's raw logit \(T_1\) through the \emph{student-side} debias correction, and obtain
\begin{equation}
    T_2' = f_S(T_1),
    \label{eq:teacher_rebias}
\end{equation}
which places the teacher prediction in the same corrected space as the student output \(S_2=f_S(S_1)\). We then perform distillation by matching \(T_2'\) and \(S_2\), rather than directly using \(T_2\). Formally, the distillation objective is defined as
\begin{equation}
    \mathcal{L}_{\mathrm{distill}}=D\!\left(T_2',\, S_2\right),
    \label{eq:kd_debias}
\end{equation}
where \(D(\cdot,\cdot)\) denotes the distillation loss. Empirically, we find that directly distilling from the teacher's own debiased output \(T_2\) to \(S_2\) leads to a substantial degradation in transfer performance, which validates the necessity of re-correcting teacher logits under the student-side debias function.

\subsection{System Design}
The model training paradigm in industrial recommendation and advertising scenarios generally comprises two phases: batch training and streaming training. The batch training phase accelerates model convergence while simultaneously boosting training efficacy. Therefore, the distillation process must maintain efficient and stable information transfer across both phases. Concurrently, it must remain mindful of the stringent training cost constraints inherent in industrial settings.

To satisfy these requirements, we have implemented a hybrid batch-streaming processing pipeline, as depicted in \figref{fig:stream}.

\subsubsection{Batch Distillation Phase:} 
In this initial phase, the objective is to rapidly converge the student model to a robust performance baseline. This phase guarantees iteration efficiency by allocating more training resources and utilizing larger batch sizes to accelerate student convergence. The distillation framework ensures that the student model can access distillation signals either by directly reading from data streams or via intermediate storage media.

\subsubsection{Streaming Distillation Phase:} 
The student model subsequently transitions into the streaming phase to process continuous influxes of new user interaction data. 
During this phase, while the data volume per step is reduced, the demand for real-time processing becomes more stringent. 
The teacher model writes the logits of the novel streaming data into low-latency, high-speed intermediate storage. Subsequently, the original data stream is joined with the distillation signals to construct a new supplementary data stream, from which the student continuously consumes the latest signals alongside their corresponding ground-truth labels. 

This empowers the student to swiftly adapt to shifts in data distributions, user behaviors, or item catalogs, ensuring that the model remains both high-performing and responsive to the dynamic nature of real-time recommendation systems. 
Creating this supplementary data stream effectively mitigates the read-amplification bottlenecks frequently encountered in industrial production when a multitude of student models directly query the high-speed intermediate storage. 
Furthermore, this newly introduced student data stream establishes an optimal trade-off between retry limits and latency, preserving its performance edge with minimal (minute-level) delay.

In industrial recommendation systems, improving iteration stability also requires stable teacher supervision across successive experiments, so that experimental conclusions can be reliably reproduced. To this end, the teacher signals consumed by the current student are typically materialized from intermediate storage or streaming data flows into long-term storage (e.g., HDFS), where they can be reused in subsequent batch training.

%% file: subfile/sec-exp.tex
\section{Experiments}
\label{sec-exp}

We conduct a comprehensive series of offline and online experiments to systematically validate the effectiveness of the proposed \modelname framework.
Our evaluation is designed to answer the following core research questions:
\begin{itemize}[topsep=0.5pt, leftmargin=*]
    \item \textbf{RQ1:} How much of the performance gain achieved by scaling up the teacher model can our distillation framework preserve in the student?
    \item \textbf{RQ2:} Can the framework effectively transfer the benefits from scaling different dimensions, including parameters, sequence length, and data?
    \item \textbf{RQ3:} Are the key components of our design necessary for achieving high distillation transferability and stable deployment?
\end{itemize}

\subsection{Experimental Setup}
We first introduce the experimental setup, including evaluation datasets, comparison methods, implementation details and evaluation settings.

\subsubsection{Datasets.}
The offline experiments in this paper are based on real-world industrial training data from multiple advertising and recommendation scenarios on the large-scale real-world short-video platforms Douyin and TikTok, including Douyin E-commerce Ads, Multi-scenario recommendation spanning products Hotsoon and the Douyin Pad client, and TikTok Live-streaming recommendation. These scenarios are representative examples, while additional scenarios are omitted for brevity.

For E-commerce Ads, the target task is Conversion Rate (CVR) prediction.
The data originates from online platform logs and user feedback labels, specifically including ad clicks, item orders, and purchase behaviors.
Each training instance comprises over one thousand features, covering numerical features, ID-based features, cross features, and long user behavior sequences whose raw lengths can exceed 20K.
The datasets span hundreds of millions of users, and after sampling, the training data scales to approximately 200 million samples per day.

For Multi-scenario recommendation, the objective is to predict multi-objective user behaviors on short-video and image-text feeds, such as completion (\textit{finish}), watch duration (\textit{staytime}), \textit{likes} and other diverse interactive objectives.
The dataset is constructed from online platform logs and user feedback labels.
The training data contains hundreds of features, including numerical features, ID-based features, cross features, and user behavior sequences, and spans hundreds of millions of users.
The total training volume scales to approximately 7 billion samples per day.

For Live-streaming recommendation, the task is to predict multi-objective user-live interactions in the ranking stage, covering both click-related and watch-engagement targets.
Following the production setting of Live-streaming, the dataset consists of real-world streaming user-item interaction data.
It contains thousands of features and dozens of prediction targets.

\subsubsection{Baselines and Implementation Details.}
The primary baseline for comparison is the \textbf{Base} model, which denotes a student-scale production model trained \textit{without} any knowledge distillation.
The exact Base and Teacher model sizes vary across scenarios.
For the Ads model setting, the Base model has 1B dense parameters and uses a 5K behavior sequence.
The \textbf{Teacher} model follows the same core architecture family but is scaled up with TokenMixer-Large and LONGER, reaching up to 24B dense parameters and 20K behavior sequence length.
We also increase the teacher's training data volume by adjusting the sampling ratio.
The \textbf{Student} model refers to the Base architecture distilled from the scaled teacher using the full \modelname framework.
All models in the same scenario are trained with the same embedding dimension and comparable core architecture family for fair comparison.
Training utilizes distributed frameworks with large batch sizes, and the hybrid batch-streaming pipeline is implemented on our internal machine learning platform.

\subsubsection{Evaluation Metrics.}
For offline evaluation, we report relative AUC gain over the corresponding non-distilled Base model.
We use the distillation transferability $\eta$ defined in \secref{sec-pre} to measure the fraction of teacher gain preserved by the student:
\begin{equation}
    \eta = \frac{P_S^{\text{distill}} - P_S^{\text{raw}}}{P_T - P_S^{\text{raw}}},
\end{equation}
where $P_T$, $P_S^{\text{raw}}$, and $P_S^{\text{distill}}$ denote the teacher, raw student, and distilled student performance, respectively.
For online A/B tests, we report business metrics from live traffic experiments.

\subsection{Distillation Gains across Scenarios}
\label{sec-exp-main}
\begin{table}[htbp]
  \centering
  \setlength{\tabcolsep}{5pt}
  \caption{Distillation gains across multiple industrial scenarios. Teacher and Student gains are measured by $\Delta$AUC against the corresponding non-distilled Base model in each scenario. Ratio denotes the student-to-teacher parameter ratio.}
  \label{tab:multi_scenario}
  \begin{tabular}{lcccc}
        \toprule
        Scenario & Teacher Gain & Student Gain & $\eta$ & Ratio \\
        \midrule
        E-com Ads & +0.69\% & +0.44\% & 64\% & 1:7 \\
        Multi-scn Rec & +0.57\% & +0.26\% & 45\% & 1:34 \\
        TikTok Live & +0.74\% & +0.37\% & 50\% & 1:5 \\
        \bottomrule
      \end{tabular}
\end{table}

\tableref{tab:multi_scenario} presents the primary results validating the \modelname framework across multiple business scenarios.
Scaling up the teacher model consistently yields substantial AUC improvements over the Base model, confirming the significant gains achievable through scaling in recommendation and advertising systems.
After distillation, the Student model preserves a large fraction of the teacher's advantage, with transferability reaching 64\% in E-commerce Ads and ranging from 45\% to 50\% in the Multi-scenario recommendation and Live-streaming scenarios.
Although the transfer rate varies across scenarios, it remains consistently high in all of them.
This demonstrates that the gains from teacher scaling can be transferred to the Student model under high compression ratios, dramatically improving serving ROI.

Notably, in the Live scenario, distillation is performed across training stages, where the teacher is the final-ranking model and the student is the pre-ranking model. Although the pre-ranking model uses fewer features and has a smaller parameter scale than the ranking model, it still retains a substantial transfer rate of 50\%. This finding further demonstrates the significance of cross-stage distillation: it enables knowledge acquired by a more expressive downstream model to be transferred to an upstream lightweight model, thereby enhancing effectiveness without sacrificing serving efficiency. This reflects that \modelname exhibits strong flexibility and generalization ability.

\subsection{Student Scaling and Transferability}
\label{sec-exp-student-scaling}

The student model is the component used for online serving.
Its own capacity and architecture therefore determine how much teacher knowledge can be absorbed and converted into real-world serving gains.
We next study the relationship between student capacity, student architecture, and transferability.

\subsubsection{Effect of Student Capacity.}
\begin{table}[htbp]
  \centering
  \caption{Effect of student capacity on distillation transferability. The teacher is fixed at 7B dense parameters.}
  \label{tab:student_capacity}
  \resizebox{\linewidth}{!}{
  \begin{tabular}{lccccccc}
    \toprule
    Student Setting & Raw Params & $\eta$ & $\Delta$Params & $\Delta\eta$ \\
    \midrule
    4 layers, dim 640 & 1B & 73\% & -- & -- \\
    4 layers, dim 320 & 376M & 65\% & -62\% & -8\% \\
    2 layers, dim 160 & 184M & 48\% & -81\% & -25\% \\
    \bottomrule
  \end{tabular}
  }
\end{table}

\begin{figure}[htbp]
    \centering
    \includegraphics[width=0.8\linewidth]{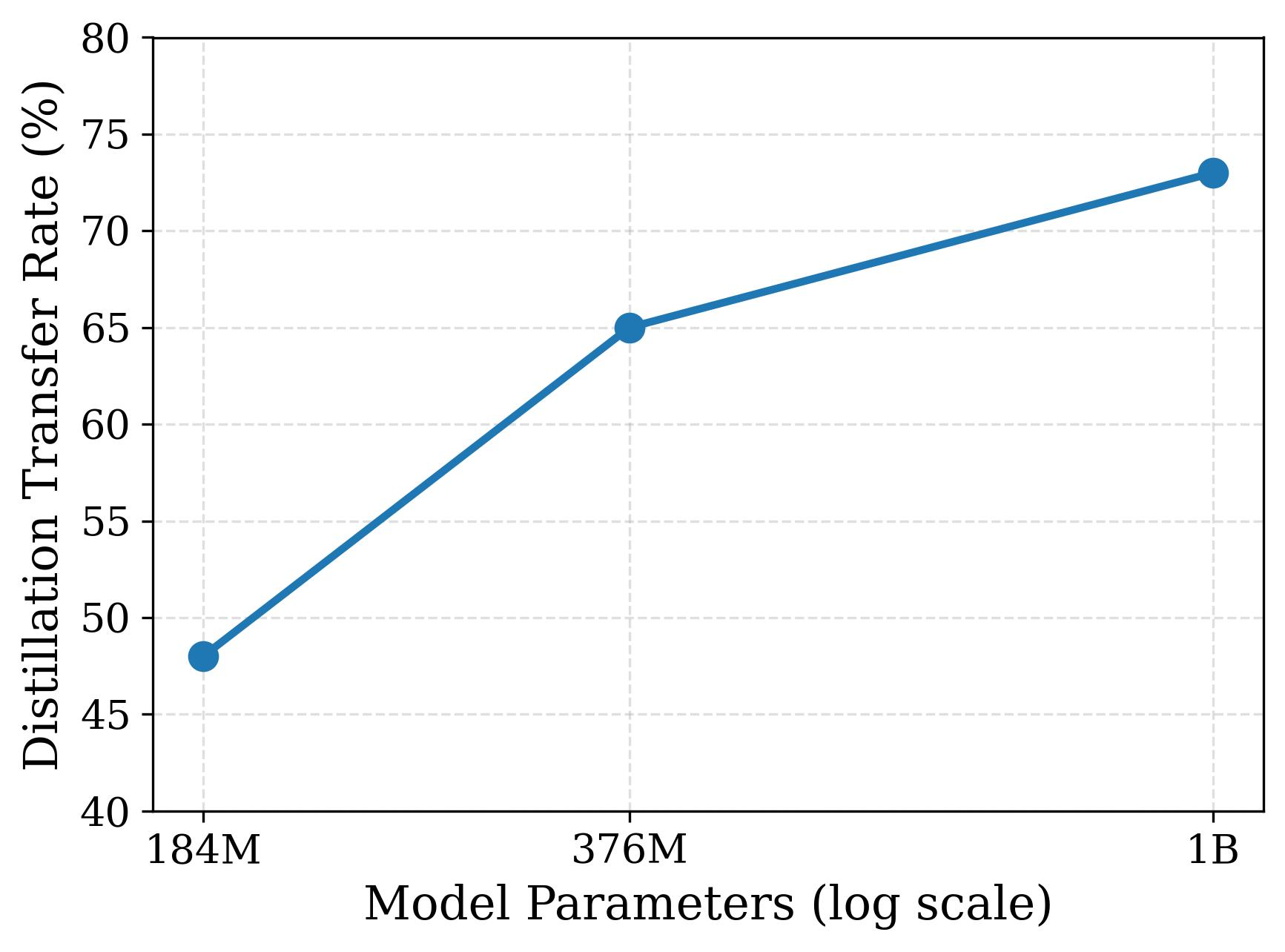}
    \caption{Distillation transferability as a function of student model capacity. Smaller students still benefit from distillation, but transferability drops as the teacher-student capacity gap increases.}
    \label{fig:student_capacity}
\end{figure}

\tableref{tab:student_capacity} and \figref{fig:student_capacity} show how student size affects distillation.
Holding the teacher and other factors constant, we vary only the student's dense parameter count.
When the teacher is fixed at 7B dense parameters, reducing the student from 1B to 376M and 184M parameters still yields stable student AUC gains, but the transferability decreases from 73\% to 65\% and 48\%, respectively.
This indicates that distillation can provide consistent improvements across different student sizes, but it cannot fully remove the transfer bottleneck caused by a large teacher-student capacity gap.
The final distillation gain is therefore jointly determined by the teacher's performance ceiling and the student's representational capacity.
When the student is too small, part of the teacher's gain becomes difficult to absorb effectively.

\begin{table}[htbp]
  \centering
  \caption{Transferability under larger teacher-student capacity gaps.}
  \label{tab:teacher_student_gap}
  \begin{tabular}{lcc}
    \toprule
    Teacher & Student & $\eta$ \\
    \midrule
    7B & 1B & 73\% \\
    24B & 1B & 65\% \\
    \bottomrule
  \end{tabular}
\end{table}

We further scale the teacher to 24B dense parameters to evaluate a more challenging capacity gap.
As shown in \tableref{tab:teacher_student_gap}, the 1B student still achieves 65\% transferability when distilled from the 24B teacher.
Although the larger gap reduces $\eta$ compared with the 7B teacher, the transferability remains high enough to convert the teacher's additional scaling gain into practical student improvements.

\subsubsection{Effect of Student Architecture.}

\tableref{tab:student_architecture} shows results for students with different depths and widths but distilled from the same large teacher.
At a similar parameter budget of about 616M, the 4-layer, 480-dimensional student achieves higher transferability than the 2-layer, 640-dimensional student.
This suggests that not only total capacity but also FLOPs or capacity allocation across depth and width influences transfer efficiency.
This result also addresses a practical production concern: under the \modelname paradigm, student iteration can still proceed through normal architecture and capacity trade-offs.

\begin{table}[htbp]
  \centering
  \caption{Effect of student depth-width allocation on distillation transferability. All students are distilled from the same large teacher.}
  \label{tab:student_architecture}
  \begin{tabular}{lcccc}
    \toprule
    Setting & Params & $\eta$ & $\Delta$Params & $\Delta\eta$ \\
    \midrule
    4 layers, dim 640 & 1B & 73\% & -- & -- \\
    4 layers, dim 480 & 616M & 68\% & -38\% & -5\% \\
    2 layers, dim 640 & 616M & 63\% & -38\% & -10\% \\
    2 layers, dim 480 & 406M & 58\% & -59\% & -15\% \\
    \bottomrule
  \end{tabular}
\end{table}

\subsubsection{Distillation Preserves the Student Scaling Law.}
A common concern is that introducing distillation may interfere with the student's own scaling behavior and thus restrict future student-side iteration.

\begin{figure}[htbp]
    \centering
    \includegraphics[width=0.8\linewidth]{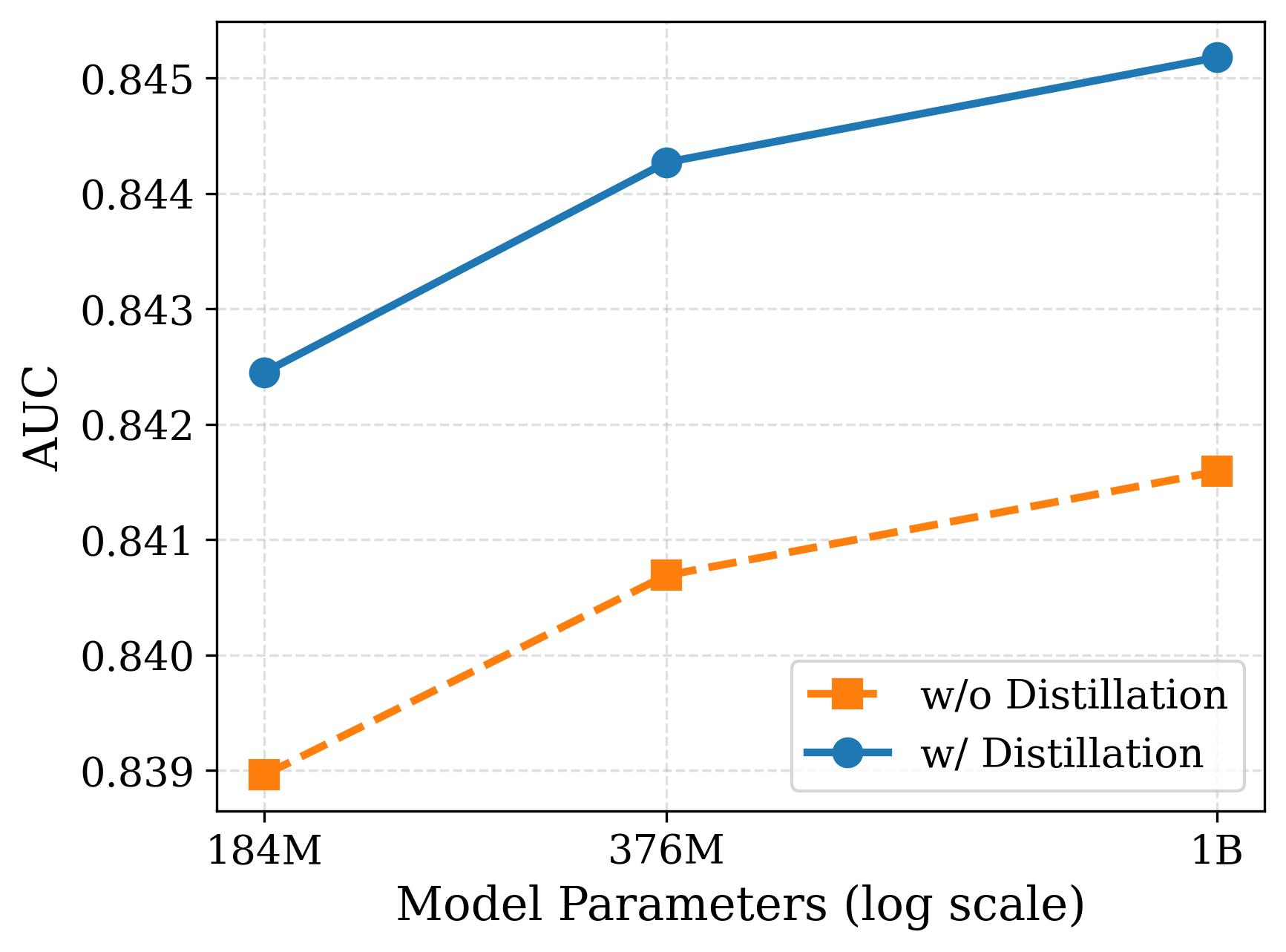}
    \caption{Student scaling trend before and after distillation. Distillation raises the overall AUC curve while preserving the relative scaling behavior of the Base model.}
    \label{fig:student_scaling_law}
\end{figure}

\figref{fig:student_scaling_law} shows that this does not occur in our setting.
As the student becomes larger, the distilled model follows nearly the same relative scaling trend as the non-distilled Base model.
Distillation therefore shifts the student scaling curve upward by providing additional supervision, rather than replacing or distorting the role of student model capacity.

\subsection{Necessity of the Decoupled-Tower Design}
\label{sec-exp-ablation}

We conduct ablation experiments to validate the necessity of the decoupled-tower structure.

\begin{table}[htbp]
    \centering
    \setlength{\tabcolsep}{4pt}
    \caption{Performance comparison of different tower structures and loss configurations.}
    \label{tab:decoupled_tower}
    \begin{tabular}{lcccc}
    \toprule
    \multirow{2}{*}{\textbf{Structure}} & \multicolumn{2}{c}{\textbf{AUC Gain}} & \multicolumn{2}{c}{\textbf{Loss Config}} \\
    \cmidrule(lr){2-3} \cmidrule(lr){4-5}
     & \textbf{Main} & \textbf{Aux} & \textbf{Main} & \textbf{Aux} \\
    \midrule
    Single & +0.03\% & - & $\mathcal{L}_{distill}$ & - \\
    Single & +0.36\% & - & $\mathcal{L}_{task} + \mathcal{L}_{distill}$ & - \\
    \midrule
    Decoupled & +0.41\% & +0.25\% & $\mathcal{L}_{task}$ & $\mathcal{L}_{distill}$ \\
    Decoupled & \textbf{+0.44\%} & \textbf{+0.38\%} & $\mathcal{L}_{task}$ & $\mathcal{L}_{task} + \mathcal{L}_{distill}$ \\
    \bottomrule
    \end{tabular}
\end{table}

\tableref{tab:decoupled_tower} presents an ablation study evaluating the effectiveness of the student model's architecture. The empirical results clearly demonstrate that the single-tower structure performs worse than our proposed decoupled-tower design. Notably, relying exclusively on distillation supervision in a single tower ($\mathcal{L}_{distill}$ only) yields a marginal AUC improvement of merely +0.03\%. 
While combining task and distillation losses within a single tower ($\mathcal{L}_{task} + \mathcal{L}_{distill}$) improves performance to +0.36\%, it still falls short of the decoupled architectures.

By explicitly decoupling the serving and distillation objectives, adding a separate main tower trained exclusively on the ground-truth task loss yields superior results (+0.41\% AUC). 
Furthermore, incorporating $\mathcal{L}_{task}$ alongside $\mathcal{L}_{distill}$ in the auxiliary tower provides an additional boost, pushing the main tower's AUC gain to an optimal +0.44\%. This enhancement likely stems from a better alignment of the feature representations learned by the shared backbone.

It is worth noting that, as discussed earlier, the ratio between $\mathcal{L}_{task}$ and $\mathcal{L}_{distill}$ applied on the auxiliary tower is highly sensitive. Improper ratio configuration will lead to a performance loss of no less than 0.05\%.

Beyond absolute performance gains, this experiment also supports the architectural advantage of the fault-tolerance design discussed earlier. 
By physically separating online serving predictions (Main Tower) from the distillation process (Auxiliary Tower), the decoupled design ensures stable and robust inference in production environments, shielding the core recommendation task from potential instabilities in the distillation pipeline.

\subsection{Teacher Scaling Ablation}
\label{sec-exp-teacher-scaling}

We analyze the contribution of different teacher-side scaling dimensions.
The results show that dense parameter scaling, sequence length scaling, and data scaling all produce teacher gains that are transferable to the student.

\subsubsection{Dense and Sequence Scaling.}
\begin{table}[htbp]
  \centering
  \caption{Ablation on individual teacher scaling dimensions. Each experiment scales one dimension while keeping the other dimensions at the baseline level.}
  \label{tab:teacher_scaling_dimension}
  \begin{tabular}{lccc}
    \toprule
    Scaling Dimension & Teacher & Student & $\eta$ \\
    \midrule
    Dense (1B $\rightarrow$ 7B) & +0.30\% & +0.22\% & 73\% \\
    Sequence (5K $\rightarrow$ 20K) & +0.19\% & +0.11\% & 61\% \\
    \bottomrule
  \end{tabular}
\end{table}

To disentangle the contribution of scaling dense parameters versus sequence length, we conduct controlled experiments where we scale only one dimension at a time.
Results in \tableref{tab:teacher_scaling_dimension} show that scaling either dimension individually provides a clear performance boost to the teacher.
More importantly, a large portion of these individual gains is preserved through distillation, with transferability rates $\eta$ of 73\% for parameter scaling and 61\% for sequence length scaling.
This confirms that our framework is effective at transferring gains from different scaling strategies, rather than depending on a single source of improvement.

\subsubsection{Data Scaling.}
\begin{table}[htbp]
  \centering
  \caption{Effect of teacher and student data scaling. Gains are measured against the non-data-scaled Base model.}
  \label{tab:data_scaling}
  \resizebox{\linewidth}{!}{
  \begin{tabular}{lccc}
    \toprule
    Setting & Teacher Gain & Student Gain & $\eta$ \\
    \midrule
    w/o data scaling & +0.42\% & +0.28\% & 67\% \\
    with data scaling & +0.69\% & +0.44\% & 64\% \\
    \bottomrule
  \end{tabular}
  }
\end{table}

We investigate whether gains from scaling the teacher's \textit{data volume} are transferable.
As shown in \tableref{tab:data_scaling}, increasing the teacher's training data volume through sampling-ratio adjustment raises teacher gain from +0.42\% to +0.69\%.
Crucially, the distilled student, which does not use the enlarged training data but learns from the data-scaled teacher, also improves from +0.28\% to +0.44\%.
This indicates that the knowledge learned by the teacher from the additional data is transferred to the student.
The student thus benefits from more data without paying the computational cost of processing it, highlighting another efficiency advantage of the framework.

\subsection{Necessity of the Hybrid Training Pipeline}
\label{sec-exp-batch-streaming}

To validate the streaming component, we compare a student updated with continuous distillation signals against one that, after batch convergence, is trained only on new data with ground truth labels and no teacher signals.

\begin{table}[htbp]
  \centering
  \caption{Ablation of streaming distillation. Performance is measured five days after batch distillation convergence.}
  \label{tab:streaming_ablation}
  \begin{tabular}{lcc}
    \toprule
    Setting & with stream & w/o stream \\
    \midrule
    $\Delta\text{AUC}$ (\%) & +0.10\% & +0.05\% \\
    \bottomrule
  \end{tabular}
\end{table}

As shown in \tableref{tab:streaming_ablation}, after five days, the student without streaming distillation retains only half of its initial gain (+0.05\% vs. +0.10\%).
This demonstrates that in a non-stationary environment, continuous distillation signals from the teacher are essential to counteract performance drift due to data distribution shifts.

\begin{figure}[htbp]
    \centering
    \includegraphics[width=0.8\linewidth]{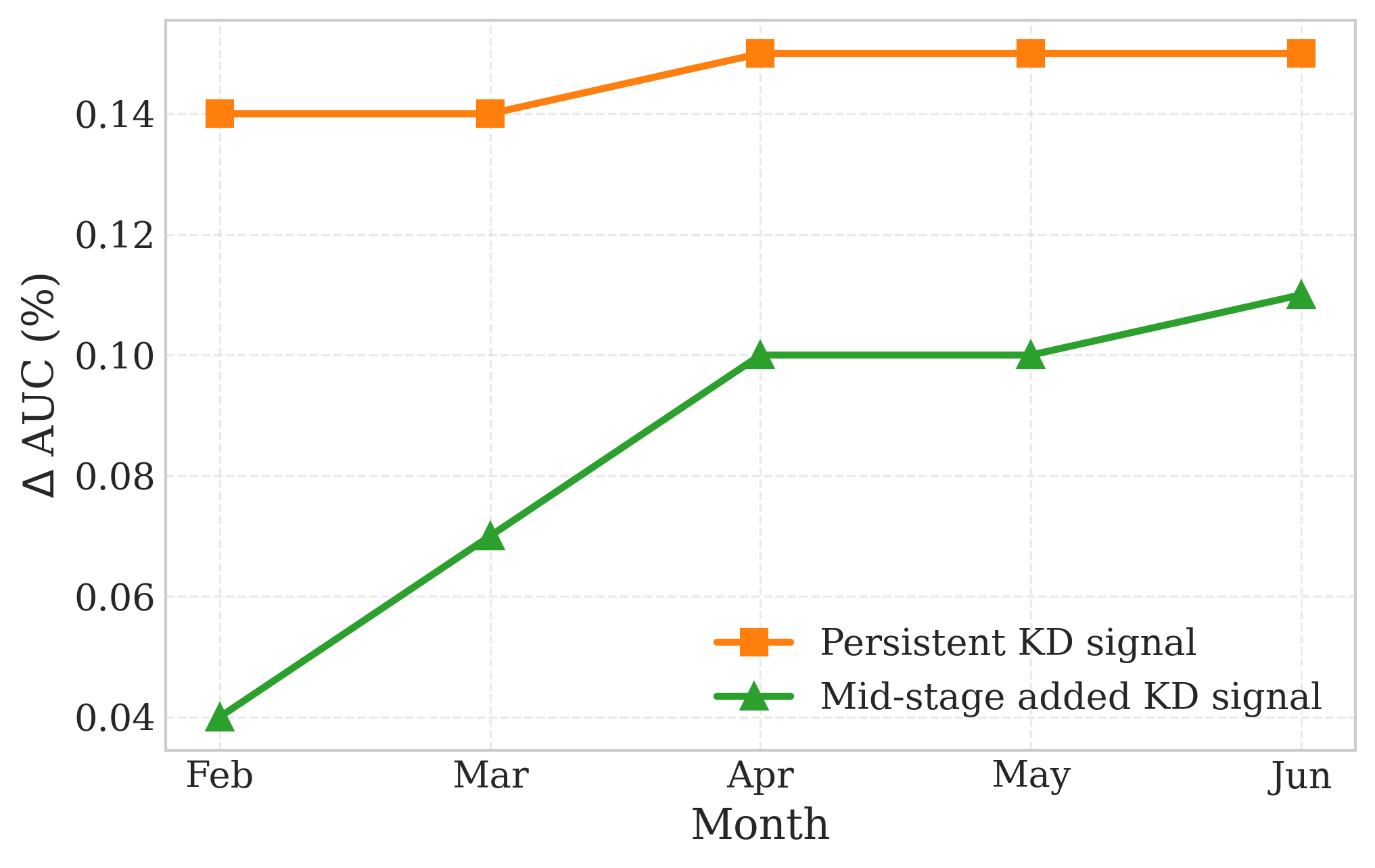}
    \caption{Ablation of the batch distillation phase. Batch distillation provides fast initialization, while streaming-only distillation from scratch converges too slowly for practical deployment.}
    \label{fig:batch_ablation}
\end{figure}

Conversely, we evaluate the importance of the batch phase by attempting to train a student using only the streaming pipeline from scratch.
Specifically, we simulate a training configuration where distillation is omitted during the batch phase and introduced only during the subsequent streaming phase.
As illustrated in \figref{fig:batch_ablation}, even after incorporating distillation signals and undergoing more than 5 months of continuous streaming training, this control group consistently underperforms the group that maintains distillation across both batch and streaming phases. This shows that initializing the student with knowledge distillation during the batch phase plays an irreplaceable role in establishing its performance ceiling.

\subsection{Necessity of the Debiasing Mechanism}
We evaluate the debiasing module designed to correct for distribution shift between teacher and student data.

\begin{table}[H]
    \centering
    \caption{Ablation study of the debiasing mechanism.}
    \begin{tabular}{lc}
    \toprule
         Method & $\Delta\text{AUC}$(\%)\\
    \midrule
         w/o debias & +0.34\% \\
         with debias & +0.44\% \\
    \bottomrule
    \end{tabular}
    \label{tab:debias}
\end{table}
As shown in \tableref{tab:debias}, removing debiasing reduces student performance from +0.44\% to +0.34\%. 
This demonstrates that distribution mismatch is a real issue that hinders distillation efficiency, and that our debiasing mechanism effectively mitigates it, contributing to the overall high transferability.

\subsection{Online A/B Test Results}
\label{sec-exp-online}

The \modelname distillation framework exhibits strong generalization and has been successfully deployed across major scenarios on the real-world short-video platforms Douyin and TikTok, including Douyin E-commerce Ads, Multi-scenario recommendation (Douyin Pad/Hotsoon), and TikTok Live-streaming recommendation. To evaluate its real-world business impact, we conduct large-scale online A/B tests across these domains, which serve hundreds of millions of active users daily. In these deployments, the distilled student models learn from a massive teacher with up to 7B dense parameters and user behavior sequences up to 20K in length. The baseline is the same student model trained without distillation.

\begin{table}[htbp]
    \centering
    \setlength{\tabcolsep}{4pt}
    \caption{Online performance improvements ($+\Delta\%$) across different scenarios. E-com Ads, Multi-scn Rec, and TikTok Live denote Douyin E-commerce Ads, Multi-scenario recommendation, and TikTok Live-streaming recommendation, respectively.}
    \label{tab:online_metrics_all}
    \begin{tabular}{llc}
        \toprule
        \textbf{Scenario} & \textbf{Metric} & \textbf{Improvement ($+\Delta\%$)} \\
        \midrule
        \multirow{4}{*}{\textbf{E-com Ads}} 
            & ADVV & +1.00\% \\
            & ADSS & +1.10\% \\
            & Order & +0.68\% \\
            & GMV & +0.62\% \\
        \midrule
        \multirow{4}{*}{\begin{tabular}{@{}l@{}}
        \textbf{Multi-scn Rec} \\
        \hspace{0.6em}(Hotsoon)
        \end{tabular}}
            & StayDuration & +0.1423\% \\
            & Play / U & +0.4677\% \\
            & Finish / U & +1.2725\% \\
            & LT & +0.0618\% \\
        \midrule
        \multirow{4}{*}{\textbf{TikTok Live}} 
            & Watch Duration / U & +0.24\% \\
            & Watch Days / U & +0.30\% \\
            & Gift Days / U & +0.28\% \\
            & Gift Revenue / U & +0.78\% \\
            \bottomrule
    \end{tabular}
\end{table}

As summarized in \tableref{tab:online_metrics_all}, the \modelname models deliver substantial and statistically significant improvements across all key metrics in every deployed scenario. In E-commerce Ads, they achieve a +1.0\% increase in ADVV and +1.1\% in ADSS, together with a +0.68\% lift in Order Count and +0.62\% in GMV. In Multi-scenario recommendation, key engagement metrics improve consistently, highlighted by a +1.2725\% boost in Finish/U and steady gains in StayDuration and Play/U. In the highly dynamic Live-streaming scenario, the \modelname models also bring notable uplifts of +0.24\% in Watch Duration and +0.78\% in Gift Revenue.

These results show that \modelname not only translates offline model gains into substantial online business value, but also generalizes well across diverse industrial scenarios. Its successful deployment across advertising, recommendation, and live-streaming further demonstrates that \modelname is a broadly applicable and scalable distillation framework for industrial recommendation systems.

%% file: subfile/sec-con.tex
\section{Conclusion}
\label{sec-con}
In this paper, we presented \modelname, a decoupled large-scale knowledge distillation framework designed to tackle the ``scale-efficiency'' bottleneck in industrial recommendation systems. 
The \modelname framework addresses the incompatibility of traditional distillation methods with recommendation scenarios, enabling high-stability and low-latency transmission of distillation signals from the teacher model to the student model with high transferability.
This is achieved by independently maximizing two critical components: the teacher model's performance gain through systematic scaling up (up to 24B dense parameters, 20K sequence length, and increased data volume), and the distillation transferability via a robust student design (a decoupled-tower architecture with debiasing), an efficient black-box distillation algorithm, and a hybrid batch-streaming processing pipeline.
Extensive offline analyses and large-scale online A/B tests across multiple core scenarios on the real-world short-video platform demonstrate that \modelname consistently achieves peak distillation transferability exceeding 60\%.
This effectively translates the capabilities of ultra-large-scale models into significant online performance gains for lightweight and cost-efficient models. 
This work not only validates the practical applicability of scaling laws in the recommendation distillation context but also provides a reusable blueprint for exploring even larger model scales in the future without incurring proportional online overhead. 
Promising directions for future work include extending the framework to multimodal recommendation tasks and further enhancing transferability in highly non-stationary environments.


